\documentclass[a4paper,
               keeplastbox,   
               ]{jacow}
\usepackage{pdfpages,multirow,ragged2e} %
\makeatletter%
	\ifboolexpr{bool{xetex}}
	 {\renewcommand{\Gin@extensions}{.pdf,%
	                    .png,.jpg,.bmp,.pict,.tif,.psd,.mac,.sga,.tga,.gif,%
	                    .eps,.ps,%
	                    }}{}
\makeatother

\ifboolexpr{bool{xetex} or bool{luatex}} 
 {}                                      
 {\usepackage[utf8]{inputenc}}           

\usepackage[USenglish]{babel}

\ifboolexpr{bool{jacowbiblatex}}%
 {%
  \addbibresource{jacow-test.bib}
  \addbibresource{biblatex-examples.bib}
 }{}
\begin{document}
\title{CHALLENGES FOR THE INTERACTION REGION DESIGN OF THE Future Circular Collider FCC-\MakeLowercase{ee}\thanks{This work was partially supported by the EC HORIZON 2020 project FCC-IS, grant agreement n.951754, and by the U. S. Department of Energy, Office of Science, under Contract No. DE-AC02-76SF-00515.}}

\author{M.~Boscolo\thanks{manuela.boscolo@cern.ch}\textsuperscript{1}, 
N.~Bacchetta\textsuperscript{2}, M.~Benedikt, H.~Burkhardt, A.~Ciarma, \\ M.~Jones, R.~Kersevan, M.~Koratzinos\textsuperscript{3}, M.~Lueckhof, K.~Oide\textsuperscript{4},\\ L.~Watrelot, F.~Zimmermann,  CERN, Geneva, Switzerland,\\ 
F.~Fransesini, L.~Pellegrino, INFN-LNF, Frascati, Italy\\
L.~Brunetti, E.~Montbarbon, M.~Serluca, IN2P3-LAPP, France, F.~Poirier, CNRS-DR17, France \\ 
M.~Dam, NBI,   Copenhagen, Denmark\\
M.~Migliorati, SBAI, Sapienza University, Rome, Italy\\
A.~Novokhatski, M.~K.~Sullivan, SLAC, Menlo Park, CA, USA\\
\textsuperscript{1} also at INFN-LNF, Frascati, Italy,
\textsuperscript{2} also at INFN-Padova, Padova, Italy,\\
\textsuperscript{3} also  at MIT, Massachusetts, USA,
\textsuperscript{4} also  at KEK, Ibaraki, Japan}

\maketitle
\begin{abstract}
The FCC-ee is a proposed future high-energy, high-intensity and high-precision lepton collider. Here, we present the latest development for the FCC-ee interaction regions, which shall ensure optimum conditions for the particle physics experiments. We discuss measures of background reduction and a revised interaction region layout including a low impedance compact beam chamber design. We also discuss the possible impact of the radiation generated in the interaction region including beamstrahlung.  
\end{abstract}

\section{IR layout}
\begin{figure}
\centering
\resizebox{0.45\textwidth}{!}{
\includegraphics{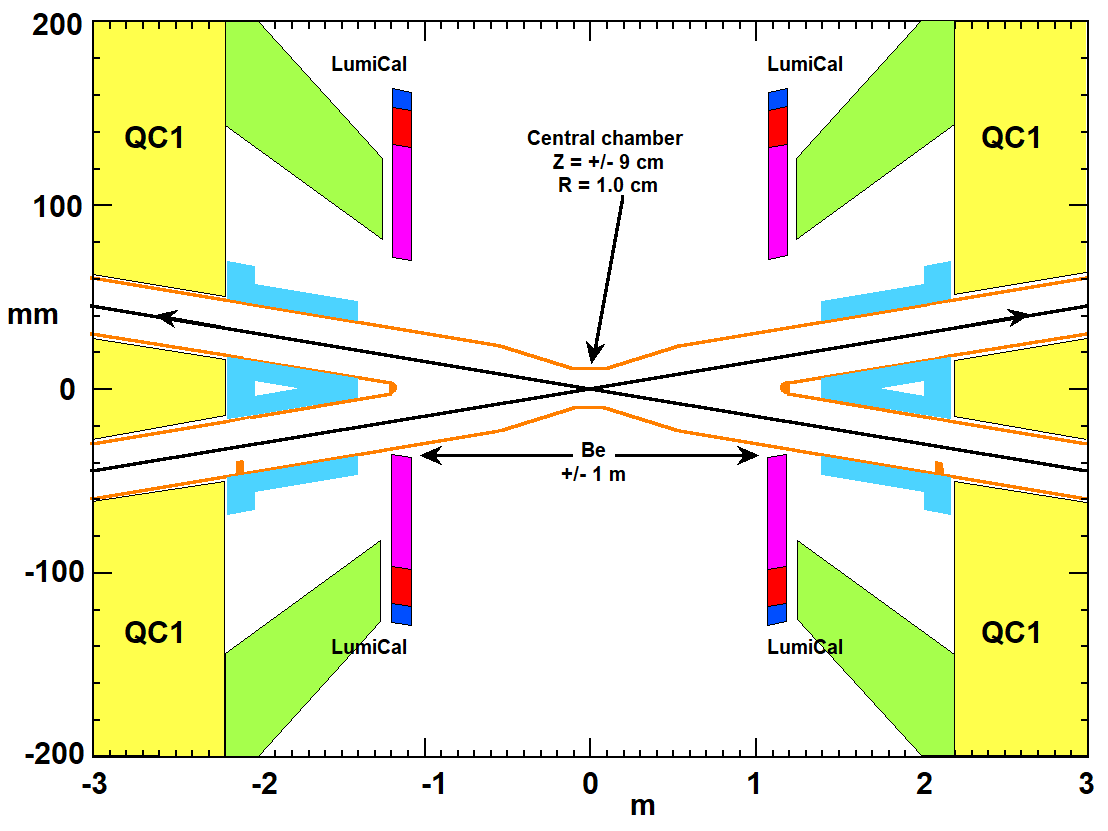}
}
\caption{IR layout with 10\,mm radius of the central pipe. }
\label{fig:1}     
\end{figure}

The $\rm e^{+}e^{-}$ collider (FCC-ee) with about 100\,km circumference has a centre of mass energy range between 90 (Z-pole) and 375\,GeV ($t\bar{t}$) and unprecedented luminosities~\cite{Abada:2019zxq}.
The flexible interaction region (IR) design is based on the crab-waist collision scheme~\cite{ref:cw} at all energies. 
 The required large crossing angle, chosen as 30\,mrad, is obtained by strongly bending the outgoing beam trajectories from the interaction point (IP) so that the beams can successfully merge back close to the opposite ring~\cite{Oide:2016mkm}, thus ensuring that most of the locally generated synchrotron radiation does not strike the IR central beam pipe.
The FCC-ee optics design maintains critical energies below 100\,keV from bending magnets 
starting from 500\,m from the IP for the incoming beam trajectories.
The IR layout is shown in Fig.~\ref{fig:1}.
One of the main constraints of the design comes from the physics requirement to keep the accelerator components within a cone below 100\,mrad from the IP  along the Z axis. This directly translates into the requirement of a challenging compact MDI design with tight space constraints.\par
The detector solenoid field is 2\,T at the IP (being a cylinder with a half-length of 4\,m and a diameter of around 3.8\,m) and it is properly compensated with two anti-solenoids.
A {\it compensation} solenoid is placed at 
1.23\,m from the IP and  a {\it screening} solenoid is placed around the final focus quadrupoles to produce an opposite field to that of the detector, thereby cancelling the detector field. Fig.~\ref{fig:1} shows  the face of the first final focus quadrupole QC1, and the {\it free length}  from the IP, ${\ell^{*}}$, of 2.2\,m.
This poses a challenge to the design of the QC1, in that the distance between the  magnetic centres of the two beams is only a few centimetres.\par
The QC1 magnets are positioned inside the detector in cantilever mode and the small beam emittances and vertical beam sizes of a few tens of nanometers at the IP call for  vibration mitigation  in this area.
Two strategies of passive and active control are usually carried out to limit the displacement of  sensitive elements.
Solutions were studied at ATF2~\cite{ATF2}. Dedicated optics and tracking simulations have started to assess the interplay of the final focus magnets, supports, and the positioning systems for the MDI area.
Beam orbit and luminosity feedbacks will be essential to control the emittance blow-up and suppress the position jitter at the IP.\par
The tight space constraints, the cantilever configuration, the difficult positions of the IR components to be measured, {\it i.e.} SC IR magnets, luminosity counter, beam position monitor,
prevent the use of current alignment and monitoring solutions. In fact,
available solutions like stretched wires require anchoring points, stable references and clear and large lines of sight that cannot be provided in our case. 
Alignment sensors, adapted to tight space constraints, must also be  radiation and cold resistant. 
Studies of alternative and innovative solutions have started.
\subsection{Low impedance beam pipe model}
\begin{figure}
\centering\includegraphics[width=0.48\textwidth]{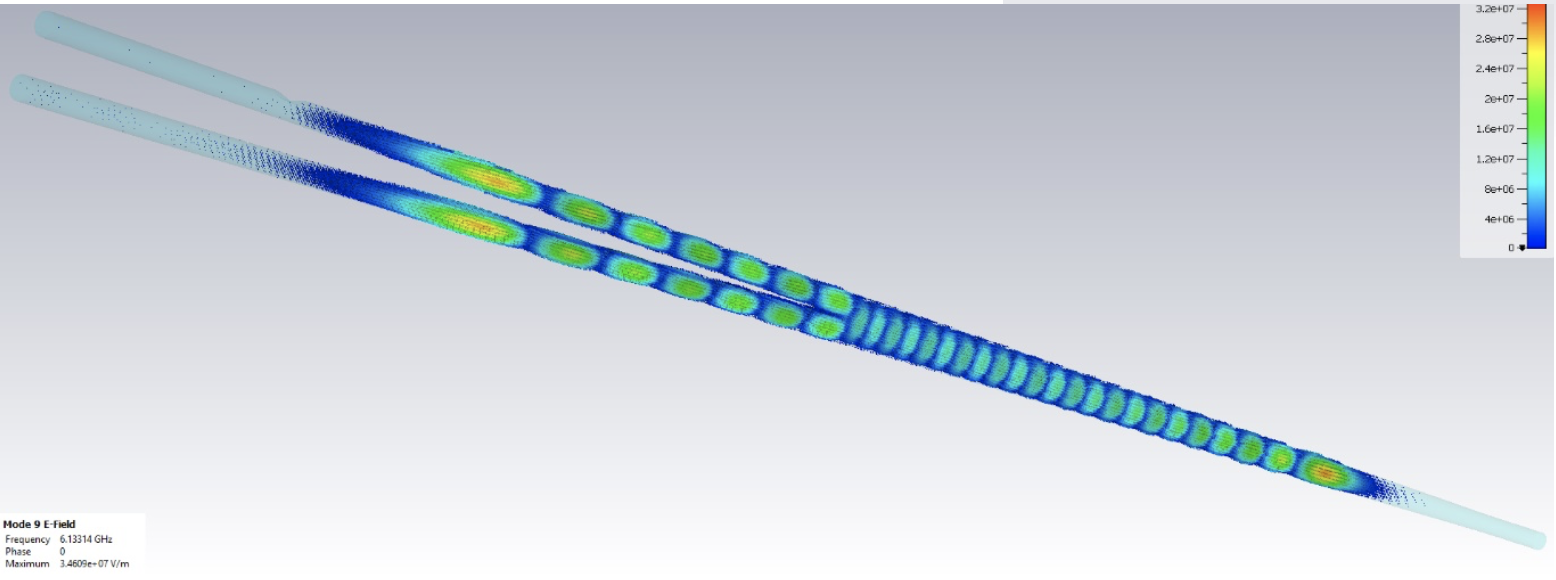}
\caption{CST simulation of the low impedance pipe model.}
\label{fig:cstbeampipe}
\end{figure}
The IR beam pipe developed for the CDR has a  15\,mm central radius for $\pm$12.5\,cm from the IP. 
We present here a novel design with a 10\,mm  central radius for $\pm$ 9\,cm from the IP  (see Fig.~\ref{fig:1}). Outside the central chamber the pipe is unchanged, and  the two symmetric beam pipes  with radius of 15\,mm are  merged together  at  1.2\,~m from the IP~\cite{novok_fccw20}.
The CAD model used for the wake field calculations has a 5\,m long beam pipe and it includes the SR masks at 2.1\,m before the IP on the inner side of the chamber centered in the radial plane and with the tip height increased from 10\,mm to 7\,mm from the beam center following the study of the SR rates described in the next two sections. Fig.~\ref{fig:cstbeampipe} shows the wake field  simulation obtained with the 3D EM field simulation software CST~\cite{cst}.\par
\begin{figure}[b]
\centering\includegraphics[width=0.38\textwidth]{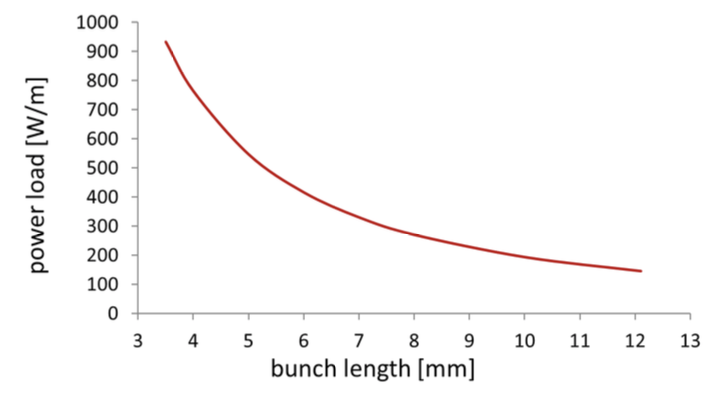}
\caption{Heat load due to resistive wakefields vs bunch length for a r=10\,mm beryllium central beam pipe.}
\label{fig:heatload}
\end{figure}
The double effect of smoothing the geometry and a smaller central pipe reduces the local heating power by a factor ten with respect to the CDR design. For a bunch length of 12\,mm, the nominal value with beams in collision at the Z-pole energy, the 
loss factor is 0.0035\,V/pC per beam, corresponding to a heating power of  260\,W for the two beams and most of this power will travel out away from the IP.
The distributed trapped modes have very small shunt impedance and even at the resonance they produce less than 200\,W for the two beams.
To remove the remaining heat, we will use a liquid coolant that will flow within the room temperature pipe. We plan to use  paraffin in the central chamber and water outside.
We are considering a light material for this central pipe called AlBemet, similar to  Beryllium.\par
With this lower impedance chamber  the higher mode absorbers (HOM)  designed to capture trapped modes and propagating waves~\cite{Novokhatski:2017kpw} are no longer foreseen.\par
The finite conductivity on the metal walls of the IR beam pipe adds an unavoidable contribution to the heat load in the IR. As is shown in Fig.~\ref{fig:heatload} the heating power strongly increases as the bunch length decreases. In our case of colliding beams at 45.6\,GeV 
$\sigma_{z}$=12\,mm the Be pipe takes 150\,W/m.
A few microns of gold coating can greatly reduce the heat load at the Z-pole, and at the  $t\bar{t}$ the gold layer helps protect the detector from SR photons.
Overall, the contribution of the IR to the total machine broadband impedance can be considered negligible.
\section{SR Backgrounds}
An efficient SR masking and shielding design used to  protect the detectors and produce a sustainable maximum occupancy has been studied and described in the CDR (see also Refs.~\cite{ref:prab_mdi, Voutsinas:2017eca, Boscolo:2019awb, Voutsinas:2020soc}).
The SR background has also been initially studied for the smaller central chamber, where it is shown how to modify the SR masks to the new design~\cite{sullivan_fccw20}. 
More detailed simulations are needed to determine the impact of the SR with this smaller chamber, SR photons have to be tracked with Geant4~\cite{geant4} in the vertex as well as in all detectors to
verify the background rates are still acceptable. A first look was encouraging, with an improvement especially at 45.6\,GeV~\cite{leogrande}, but more detailed studies are in progress and will help in finding the best design. We note that the smaller beam pipe will also increase the background rate from beam-gas events, and we have not completed this study yet.\par
To reduce the SR backgrounds to tolerable limits in addition to a proper IR optics design, a combination of fixed and movable masks, collimators and shielding have been designed.
Three collimators upstream the last bending magnet from the IP, to be sloped so that they are unable to forward scatter photons toward the IP, are proposed in addition to the fixed mask  at 2.1\,m before the IP.
With the 15\,mm radius pipe considered for the CDR study there are no direct hits of SR photons on the entire central beam pipe for $\pm$1~m around the IP.
The SR background comes from the scattering of the SR photons from the tip of the masks located at $-$2.1\,m on the incoming sides of the IP beam pipes. This is true for all beam energies (45.6, 80, 125 and 182.5\,GeV).
These masks are 5\,mm high, corresponding to 10\,mm from the beam center line. 
Only at the highest energy the scattered $t\bar{t}$ SR photon energies are high enough to penetrate more of the mask tip and the nearby beam pipe wall.\par
With the smaller pipe the SR fan from the last  bending magnet at $-$100\,m now hits the tapered upstream part of the pipe,
from $-$40 to $-$9\,cm with about 9\,W of SR power~\cite{sullivan_fccw20}.
This surface has a view to the central chamber and some of the incident photons will forward scatter to the central part of the Be chamber as well as directly penetrate the Be wall they hit.
\begin{figure}
\centering
\resizebox{0.45\textwidth}{!}{
\includegraphics{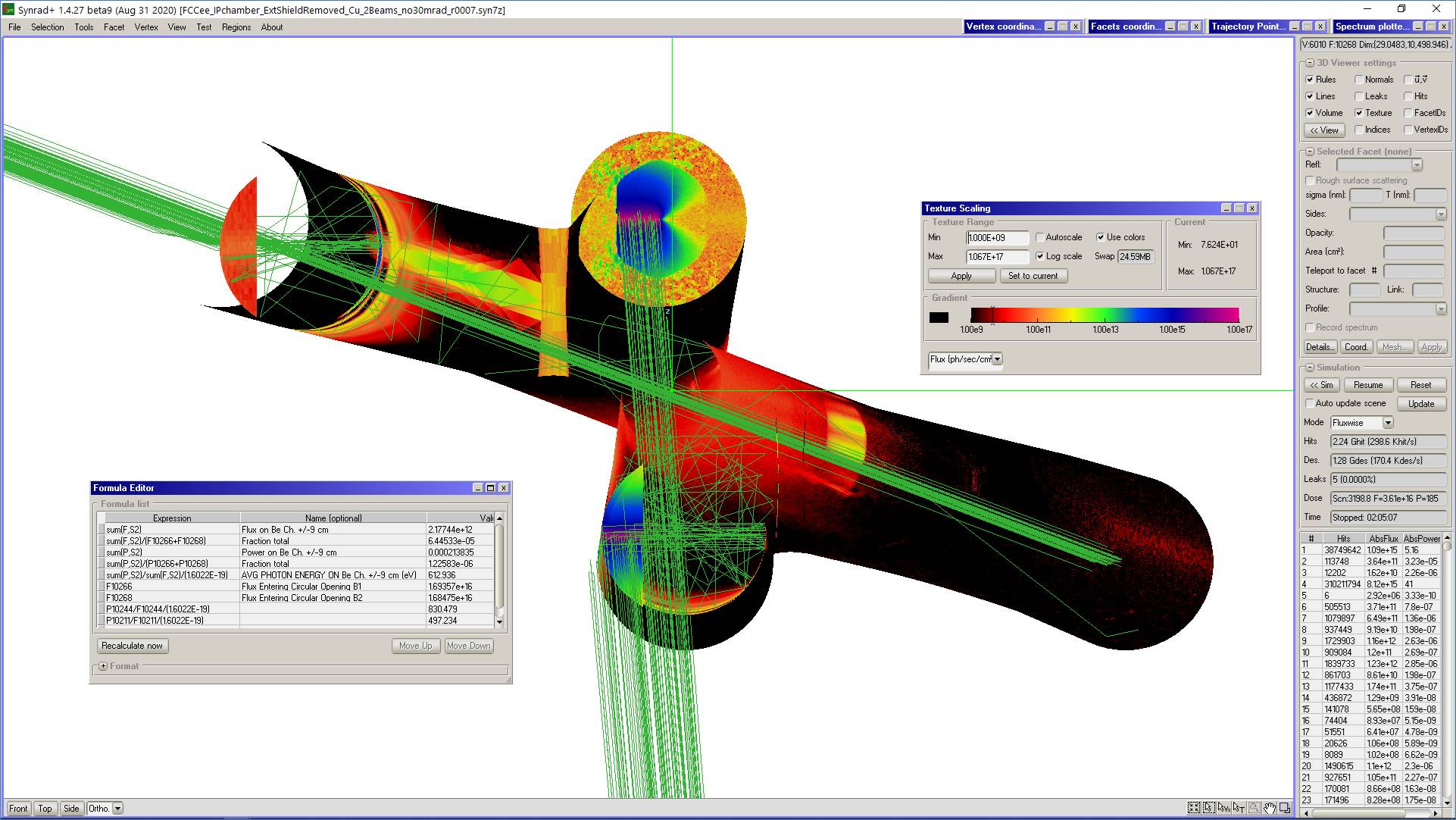}
}
\caption{Synrad+ simulation from ~120\,m toward the IR with the new smaller central Be pipe with 10\,mm radius.}
\label{fig:sr}     
\end{figure}
This direct hit rate for the Z is acceptable, as the photon energies are quite low.
Further inserting the mask tip at $-$2.1\,m from 10 to 7\,mm from the beam center the central chamber is once again shadowed by this mask tip against direct hit SR.\par
A ray-tracing simulation carried out with the SYNRAD+ code~\cite{ref:synrad} is shown in Fig.~\ref{fig:sr}. The two incoming SR beams are shown in green. The two half-moon shaped masks, placed inside the corresponding pipes, prevent the primary SR photons generated by the two last dipoles, ~ 100\,m away from the IP, from directly hitting the beryllium pipe in the center. The masks  are also sketched in Fig.~\ref{fig:1}. Only a very small fraction, less than 0.01\% of the incoming photon flux, is scattered many times and ultimately ends up on the Be pipe. Their average energy is less than 1\,keV, negligible from the point of view of radiation damage to the vertex detector layers placed around the Be chamber. The simulation refers to the 182.5\,GeV beam energy case, where the critical energy of the last dipoles is ~100\,keV.
In addition, a small amount of the SR produced by the final focus quadrupole is now striking the 
downstream part of the Be chamber. This photon rate is low at 45.6\,GeV, and it increases with the beam energy, along with the photon energies. At 120\,GeV these photon rates seem acceptable but a Geant4 simulation through the detector is needed to verify the  actual detector response.  This Geant4 simulation  is even more necessary at $t\bar{t}$, where photon energies are higher still. 
The main SR background coming from the final focus quadrupoles depends on the particle distribution in the non-Gaussian beam tails, which depend on the stored beam conditions. In the early running of an accelerator, the beam particles that populate these tails come mostly from beam-gas scattering interactions that occur around the entire ring. As the ring vacuum improves, the non-Gaussian tails can become populated by several mechanisms: intra-bunch scattering (Touschek), coupled bunch interactions (TMCI), and the collision interactions (tune-shift, radiative Bhabhas) to name a few. The beam-tails are hence difficult to model without a full and complete understanding of the accelerator condition and status at any given time.\par
\section{Radiation generated at the IP}
\begin{figure}[h]
\centering\includegraphics[width=0.48\textwidth]{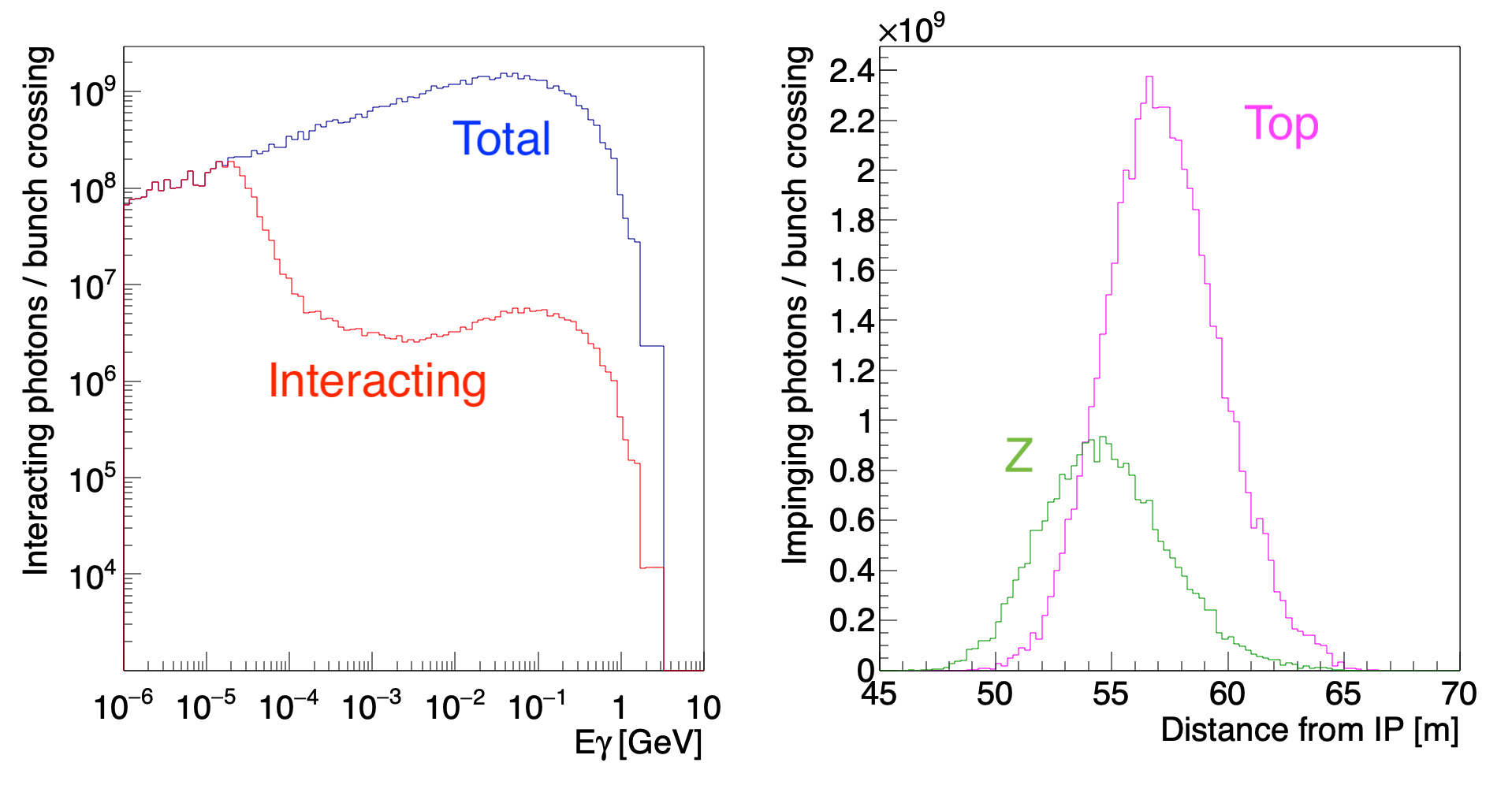}
\caption{Beamstrahlung photons: (left) energy spectra at production (blue) and  after the interaction with 1\,mm Cu pipe, for the $t\bar{t}$ case; (right) hitting location on the pipe.}
\label{fig:beamstrah2}
\end{figure}
A significant flux of photons is generated at the IP in the very forward direction by  Beamstrahlung, radiative Bhabha, and solenoidal and quadrupolar magnetic fields.  
Beamstrahlung interactions produce an intense source of locally lost beam power, as reported in the second column of Table~\ref{table_bspow}, with an average photon energy of 2 and 67\,MeV, at Z and $t\bar{t}$, respectively. 
At 45.6\,GeV the SR from IR solenoids and quads is evaluated on the order of ~37\,kW and ~7\,kW respectively with an average photon energy of ~20\,keV. For the radiative Bhabha the SR is ~0.7\,kW and $\rm <E_{\gamma}>\sim$ 5\,GeV.\par
A further study on Beamstrahlung has been performed to estimate the interacting power when hitting the pipe. Photons generated by GuineaPig++~\cite{guineapig} at  the Z and $t\bar{t}$ have been tracked from the IP to the longitudinal location 
where they hit the pipe (Fig.~\ref{fig:beamstrah2}).
The impinging angle of the photons with the pipe is about 1\,mrad for both beam energies.
The interaction probability has been evaluated for a copper beam pipe with a 1\,mm thickness.
The energy spectra of the produced and absorbed Beamstrahlung photons at the $t\bar{t}$ are shown in Fig.~\ref{fig:beamstrah2}.
Only up to about 0.5\% of the total emitted power is deposited on the pipe (last column in Table~\ref{table_bspow}).
  A dedicated study with full simulation  is needed to assess how to deal with such a high power 
 in the system in such a small space.
\begin{table}[h]
	\setlength\tabcolsep{3.5pt}
\centering
\caption{Radiation power from Beamstrahlung photons.}
\begin{tabular}{lcc}\hline
		\toprule
\textbf{Beam energy} & \textbf{Radiation Power} & \textbf{Interacting Power} \\
		\midrule
45.6\,GeV & 387\,kW & 1.8\,kW\\
182.5\,GeV & 89\,kW & 0.4\,kW\\
		\bottomrule  
\end{tabular}
\label{table_bspow}
\end{table}

\section{CONCLUSION}
We have described some of the key issues related to the FCC-ee MDI design study. We presented the recent  studies for 
a smaller and improved design of the central Be chamber.
The first look at the detector performance is encouraging, more detailed studies are in progress to check  the vertex performance and the effect of the backgrounds on other aspects of the detector performance.


\bibliographystyle{unsrt}

\end{document}